\documentclass[aps,prl,twocolumn,letterpaper,superscriptaddress,showpacs]{revtex4}
\usepackage{graphicx}
\usepackage{CJK}

\begin{document}
\begin{CJK*}{UTF8}{bsmi}
\title{
Doping effects of Se vacancies in monolayer FeSe
}
\author{Tom Berlijn}
\affiliation{Department of Physics, University of Florida, Gainesville, Florida 32611, USA}
\affiliation{Quantum Theory Project, University of Florida, Gainesville, Florida 32611, USA}
\affiliation{Center for Nanophase Materials Sciences and Computer Science and Mathematics Division, Oak Ridge National Laboratory, Oak Ridge, TN 37831-6494, USA}
\author{Hai-Ping Cheng}
\affiliation{Department of Physics, University of Florida, Gainesville, Florida 32611, USA}
\affiliation{Quantum Theory Project, University of Florida, Gainesville, Florida 32611, USA}
\author{P. J. Hirschfeld}
\affiliation{Department of Physics, University of Florida, Gainesville, Florida 32611, USA}
\author{Wei Ku(%
顧威)
}
\affiliation{Condensed Matter Physics and Materials Science Department,
Brookhaven National Laboratory, Upton, New York 11973, USA}
\affiliation{Physics Department, State University of New York, Stony Brook,
New York 11790, USA}

\date{\today}

\begin{abstract}
Following the discovery of the potentially very high temperature superconductivity in monolayer FeSe we investigate the doping effect
of Se vacancies in these materials. 
We find that Se vacancies pull a vacancy centered orbital below the Fermi energy that absorbs most of the doped electrons. 
Furthermore we find that the disorder induced broadening causes an effective hole doping. 
The surprising net result is that in terms of the Fe-$d$ bands Se vacancies behave like hole dopants rather than electron dopants. 
Our results exclude Se vacancies as the origin of the large electron pockets measured by angle resolved photoemission spectroscopy.  
Furthermore the unexpected doping effects not only lead to numerous consequences for the debated role of anion vacancies in the iron-based superconductors, 
but also demonstrate the surprising rich physics of vacancies in materials in general.
\end{abstract}


\maketitle
\end{CJK*}

The recent discovery~\cite{qywang} of monolayer FeSe grown on a SrTiO$_3$ substrate is among the most significant in the field of Fe based superconductors.
In general FeSe films play an important role for spectroscopies as pure FeSe is hard to grow as single crystals and hard to cleave along the Fe plane~\cite{sytan,rhyuan}.
By employing a Se etching technique, in which the SrTiO$_3$ substrate is smoothened by bombarding it with Se, a FeSe film was grown to be as thin as a single monolayer\cite{qywang}.  
A subsequent annealing procedure was furthermore found by scanning tunneling spectroscopy (STS) to induce a superconducting gap that is roughly an order of magnitude larger than in bulk FeSe~\cite{clsong}. The large gap was also confirmed with  ARPES \cite{dfliu, slhe,sytan}.
Direct measurement of the transition temperature $T_c$ via transport remains difficult due to the large conductivity of the SrTiO$_3$ substrate induced by the Se etching treatment~\cite{qywang}. 
Nonetheless, a crude estimate based on the gap value led to the exciting conclusion~\cite{qywang} that the $T_c$ in the FeSe monolayer could be as high as the liquid nitrogen boiling temperature of 77K.

From a fundamental science point of view, the FeSe monolayer also has the potential of being a model system for the Fe based superconductors.
Its two dimensional structure (without the complication of phase separation like in K$_x$Fe$_{2-y}$Se$_2$~\cite{zwang, ricci}) could simplify the theoretical and experimental analysis and help to identify the essential tuning parameters of the superconductivity.
One point of view ~\cite{qywang} is that the high dielectric constant of the SrTiO$_3$ substrate plays a crucial role. This is based on the fact that no dramatic $T_c$ increase is observed when the much less polarizible bilayer graphene substrate is employed ~\cite{clsong}. 
A combined functional renormalization group and Eliashberg calculation~\cite{yyxiang} indeed indicated that the dielectric phonons in SrTiO$_3$ can enhance the $T_c$ by screening the repulsive interactions in the FeSe monolayer. 
On the other hand the significantly enlarged in-plane lattice constant of the monolayer caused by the SrTiO$_3$ substrate is argued ~\cite{sytan} to be responsible for the high $T_c$ in the FeSe monolayer. Of course yet another essential tuning parameter to consider is the doping. 
 
While FeSe is expected to be a compensated semimetal, ARPES measurements ~\cite{dfliu} surprisingly revealed a strongly electron doped Fermi surface,
consisting only of electron pockets, reminiscent of the Fermi surface of K$_x$Fe$_{2-y}$Se$_2$ ~\cite{qian, zhang, mou}.
More interestingly it was shown ~\cite{slhe} that the same annealing procedure which induces the superconductivity is also responsible for the strong electron doping.
By systematically performing ARPES measurements in different stages of the annealing it was found that the hole pocket systematically vanishes~\cite{slhe}. 
To  explain the origin of the post-annealing induced electron doping two scenarios have been proposed~\cite{slhe}.
Either the electron doping is induced by Se vacancies in the FeSe monolayer or by oxygen vacancies in the SrTiO$_3$ substrate. 
In previous density functional theory (DFT) studies, the effects of oxygen vacancies~\cite{kliu,jbang,hycao}, magnetism~\cite{kliu,bazhirov, fwzheng,hycao} and electric field ~\cite{fwzheng} have been considered.
On the other hand there have been numerous reports on the occurrence of Se vacancies in iron selenides ~\cite{fchsu, margadonna,clsong,wli,mcqueen, nitsche}.
Given their high volatility, Se anions could easily evaporate during the post-annealing procedure. 
Such Se vacancies should be expected to electron dope the Fe bands strongly as they are located in the FeSe plane.
For the same reason, they should also strongly scatter the Fe-$d$ carriers. 
This is highly relevant because disorder itself can induce large effective dopings ~\cite{kxfeyse2,tm122,haverkort}.
It is therefore timely to study the influence of disordered Se vacancies on the FeSe monolayer.

In this  Rapid Communication we investigate from first principles the doping effect of disordered Se vacancies in monolayer FeSe.
We find that Se vacancies pull a vacancy centered orbital consisting of a symmetric superposition of the surrounding Fe-$s$ orbitals below the Fermi energy that absorbs most of the doped electrons. 
Furthermore we find that the disorder induced broadening induces an effective hole doping. 
The surprising net result is that Se vacancies in terms of the Fe-$d$ bands behave like hole dopants rather than electron dopants. 
Our results rule out Se vacancies to be the origin of the large electron pockets measured by ARPES.
The counter-intuitive doping effects found in our study also give new insight in the debated role of anion vacancies in the iron based superconductors, and illustrate the rich behavior of vacancies in materials in general.
 
The band structure of a disordered system is given by the configuration-averaged spectral function:
$\langle A_n(k,\omega)\rangle$=$\sum_{c}A^{c}_n(k,\omega)$ of Wannier orbital~\cite{wannier} $n$, crystal momentum $k$ and frequency $\omega$,
in which an equal probability is assumed among the configurations.
By treating the disordered configurations within the supercell approximation, their spectral functions $A^{c}_n(k,\omega)$
can be obtained directly from the supercell eigenvectors and eigenvalues, using the unfolding method~\cite{unfolding}.
To handle the computational cost related to the large sizes of the supercells, essential for a proper treatment of the disorder, we employ the recently developed Wannier function based effective Hamiltonian method~\cite{naxco2}.
The low energy Hilbert space is taken in the range [-7,3]eV consisting of the Wannier orbitals of Fe-$d$ and Se-$p$ characters.
Furthermore, a vacancy centered orbital of symmetric superposition of the surrounding Fe-$s$ orbitals is included. 
The influence of the Se vacancies is extracted from two DFT~\cite{sup} calculations: the clean Fe$_2$Se$_2$ and the single vacancy supercell
Fe$_{8}$Se$_{7}$.
The SrTiO$_3$ substrate is not included in the DFT calculations since it was found~\cite{kliu} to have no significant influence on the bands near the Fermi surface.  
To mimic the annealing-induced evaporation of the Se atoms only Se vacancies on one side of the Fe plane are considered. 
For the configurational average, we use 10 large supercells (e.g. Fig.\ref{fig:fig2}(a)) of random size, shape and orientation containing 150 atoms on average. 
 Benchmarks demonstrating the accuracy of the effective Hamiltonian method and the convergence of the configuration average against the size and the number of configurations are given in Ref.~\cite{sup}.
Atomic images are produced by the XCRYSDEN program ~\cite{kokalj}.
 
\begin{figure}[htp]
\includegraphics[width=1\columnwidth,clip=true]{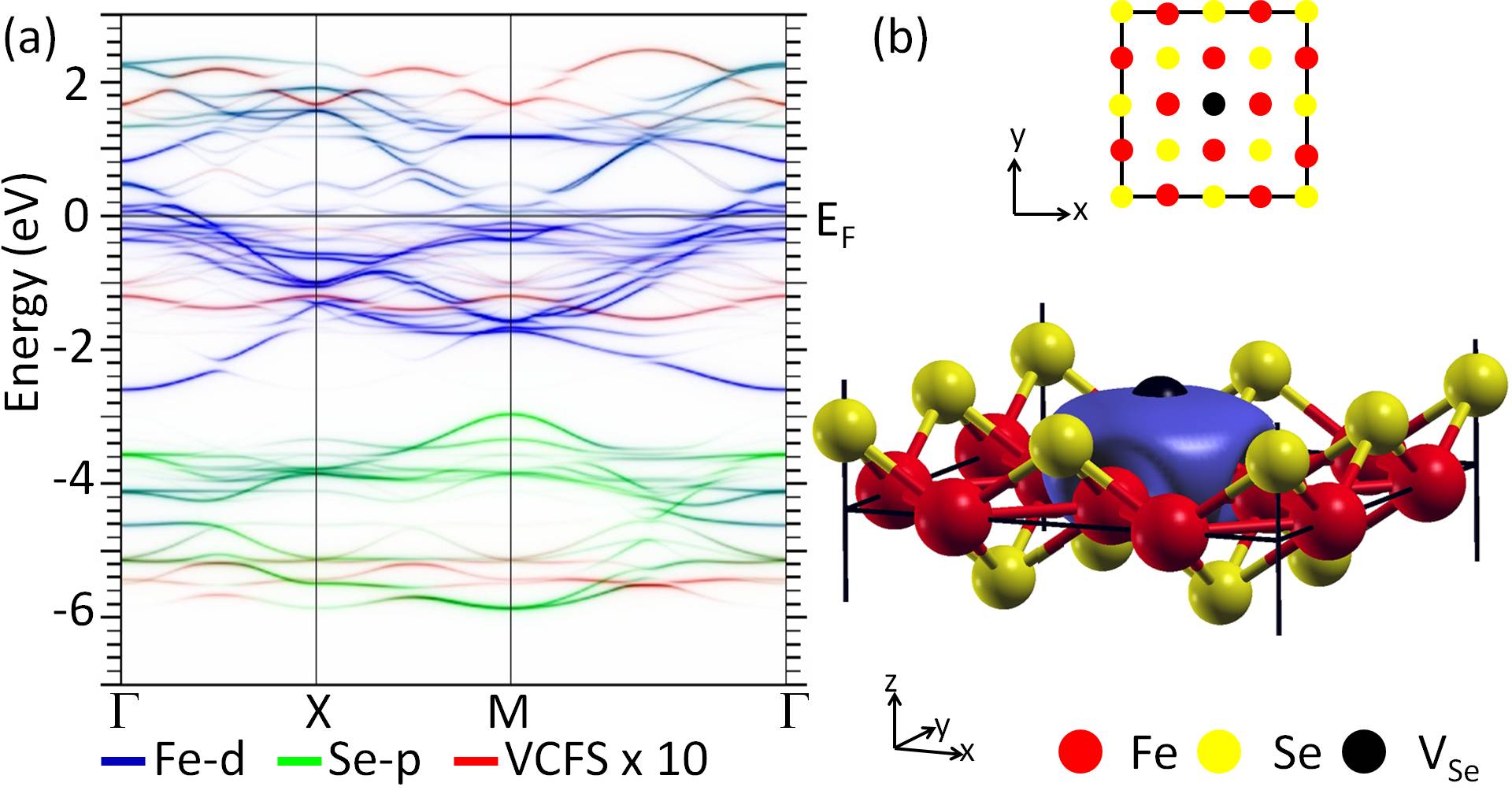}
\caption{\label{fig:fig1}
(color online) 
(a) unfolded bandstructure of the Fe$_8$Se$_7$ monolayer supercell with the vacancy centered Fe-$s$ (VCFS) intensity enhanced by a factor of 10. 
(b) isosurface (0.05 bohr$^{-3/2}$) of the VCFS orbital.
}\end{figure} 
 
We start by considering the influence of a single Se vacancy.
For this purpose, we analyze the bandstructure of the small Fe$_{8}$Se$_{7}$ supercell shown in Fig. \ref{fig:fig1}. 
The first thing we note is that there are 41 bands within the [-3,3]eV range consisting of $5\times 8$ Fe-$d$ bands leaving one band unaccounted for.
Apparently the Se vacancy pulls an extra band into the Fe-$d$ band complex.
By applying the projected Wannier function method ~\cite{sup,wannier} we find that the extra band has the character of a bonding molecular orbital formed by the four Fe-$s$ orbitals that surround the Se vacancy.
We refer to this molecular orbital as the vacancy centered Fe-$s$ (VCFS) orbital. 
From the VCFS Wannier function plot in Fig. \ref{fig:fig1}(b) we see that most of its weight is located below the Se vacancy site where the four neighboring Fe-$s$ orbitals interfere constructively. 
So why would a Se vacancy pull this Fe-$s$ like orbital down into the Fe-$d$ band complex?
After all, the Fe-$s$ orbitals in FeSe, like many outer shell transition metal $s$ orbitals in transition metal chalcogenides, are unoccupied and energetically far above the Fermi energy. 
The reason is that the Se$^{2-}$ ions in FeSe are repulsive and therefore that effectively Se vacancies will be attractive.
Of all the orbitals in the system the VCFS orbital will feel this attractive potential  most strongly as its wave function is centered right below the Se vacancy.
Another way to understand the presence of the Fe-$s$ orbital is that without the neighboring Se ligand the Fe ions that surround the Se vacancy become more like atomic Fe, which has both its $3d$ and $4s$ orbitals filled.  

\begin{figure*}[htp]
\centering
\includegraphics[width=2\columnwidth,clip=true]{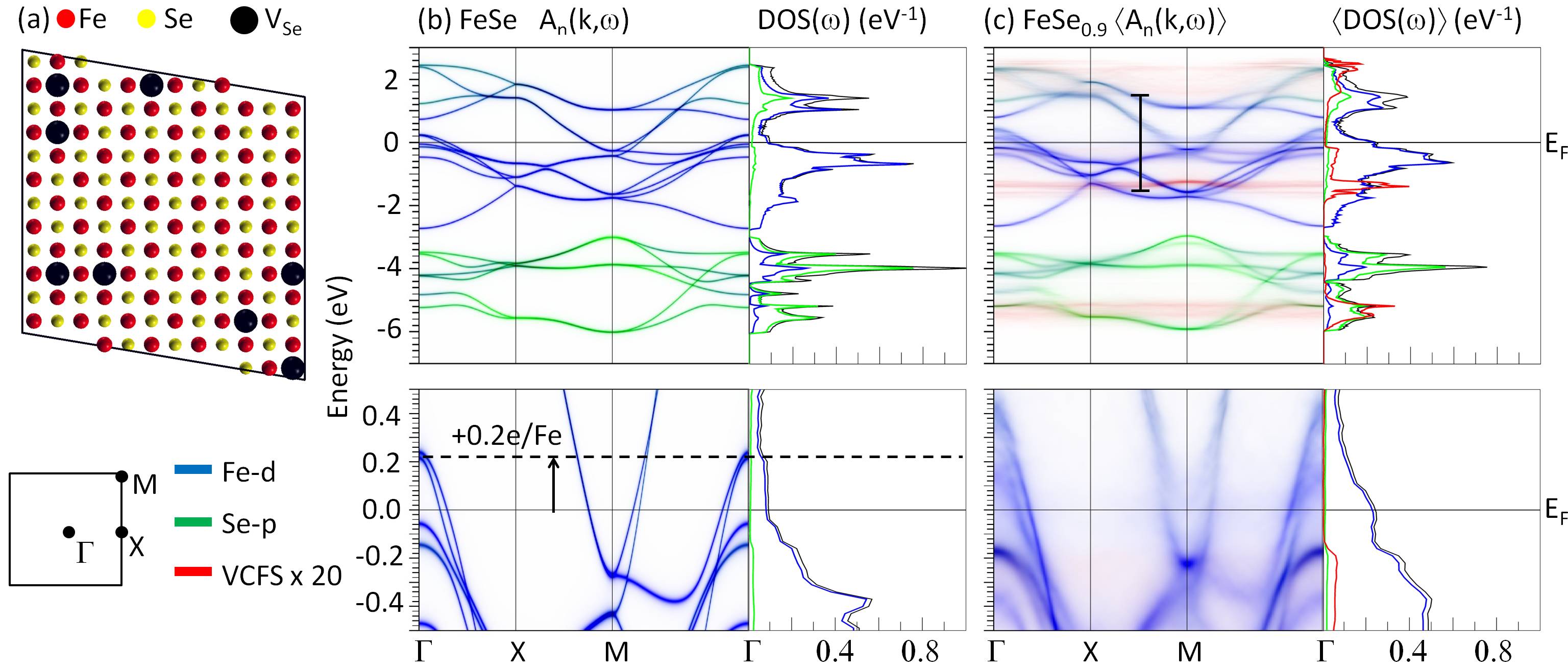}
\caption{
(color online)
(a) An example of a large sized supercell used for the configurational average.
Bandstructure and density of states of (b) clean FeSe monolayer and (c) disordered FeSe$_{0.9}$ monolayer with the VCFS intensity enhanced by a factor of 20. 
}
\label{fig:fig2}
\end{figure*}

Now let us go back to our main question whether Se vacancies can electron dope FeSe.  
If the VCFS orbital was pulled completely below the Fermi energy, then the two electrons donated by the Se vacancy would be completely absorbed.
However if we look at the unfolded band structure in Fig. \ref{fig:fig1}(a) we see that the VCFS band (indicated with the red color) is only partly occupied. 
From integrating the VCFS orbital density of states up to the Fermi energy we find that the VCFS band is filled with $\sim1.4$ electrons.
The reason for this fractional filling is the large splittings induced by the very strong hybridization ($\sim$eV) with the surrounding Se-$p$ orbitals.
This causes the VCFS spectral weight to be partly below and partly above the Fermi energy.  
The same analysis has been performed for other monolayer and bulk supercells ~\cite{sup} and for all of them the same conclusion was reached that each Se vacancy pulls down a VCFS band into the Fe-$d$ band complex with a filling of $\sim1.4$ electrons. In particular we checked~\cite{sup} that the Se vacancy induced ferrimagnetism and lattice relaxation ~\cite{kwlee} does not alter the filling of the VCFS orbital. 
The VCFS shows similar large occupancies ($\sim1.6$ electrons per As vacancy) in the Ba$_2$Fe$_4$As$_3$ and Li$_8$Fe$_8$As$_7$ supercells which demonstrates its importance for all iron based superconducting families in which anion deficiencies frequently occur. 
At this point we conclude from a single impurity point of view that unexpectedly the Se vacancy only dopes $\sim0.6$ electrons in the Fe-$d$ bands as the other $\sim1.4$ have been absorbed by the VCFS orbital.

In addition, as was demonstrated in Ref. ~\cite{tm122,haverkort,kxfeyse2}, multiple disordered impurities can conspire together to induce an effective doping in which the Fermi energy shifts without the addition or removal of physical electrons.   
Therefore let us now study the bandstructure of disordered FeSe$_{0.9}$.
As a reference we first show the bandstructure of the clean FeSe without Se vacancies in Fig. \ref{fig:fig2}(b).
In the bottom panel of Fig. \ref{fig:fig2}(b) a rigid band shift is indicated corresponding to 0.2 electrons per Fe, which is what would have naively been expected for 10 percent of Se vacancies. 
As can be seen in Fig. \ref{fig:fig2}(b) such a shift (indicated by the dotted line and the arrow) would have almost completely removed the hole pocket around $\Gamma$ from the Fermi surface. 
Such a trend would have been in good agreement with the ARPES experiments ~\cite{dfliu, slhe, sytan}, which measure no hole pocket for the annealed FeSe monolayer.

Yet it turns out that the Se vacancies behave like hole dopants rather than electron dopants.
In Fig. \ref{fig:fig2}(c) the bandstructure of the monolayer FeSe with 10 percent disordered Se vacancies is shown. 
If we compare the bands around the Fermi surface of the clean FeSe in the bottom of Fig. \ref{fig:fig2}(b) with those of the disordered FeSe$_{0.9}$ in the bottom of Fig. \ref{fig:fig2}(c) we see that the hole pockets are enhanced rather than diminished.
In particular the Fermi momentum of the outer hole sheet is enhanced by $\sim$5\% and an additional inner hole sheet emerges. 
This is in drastic disagreement with the rigid band shift in the bottom of Fig. \ref{fig:fig2}(b) in which the hole pockets were almost removed.
To a large extent, the occupied flat VCFS spectral weight shown in the top of Fig. \ref{fig:fig2}(c) is responsible for the difference.
Just as in the single impurity supercell case (Fig. \ref{fig:fig1}) the VCFS orbitals are filled with roughly $\sim$1.4 electrons. 
Still, given that the Se vacancies behave like hole dopants this leaves more than $\sim$0.6 electrons per Se vacancy unaccounted for.  

We further observe that the bands of the disordered FeSe$_{0.9}$ broaden significantly. 
These broadenings in both frequency and momentum space correspond to the finite lifetime caused by the scattering from the Se vacancies.
The fact that not only the Se-$p$ bands but also the Fe-$d$ bands broaden is important and was missed in an earlier calculation~\cite{ppsingh} based on the coherent potential approximation.  
The broadening of the Fe-$d$ bands is caused indirectly by the Fe-$d$/Se-$p$ hybridization and directly by the strong but non-local impurity potential of the Se vacancies that act on the surrounding Fe-$d$ orbitals.

\begin{figure}[htp]
\includegraphics[width=1\columnwidth,clip=true]{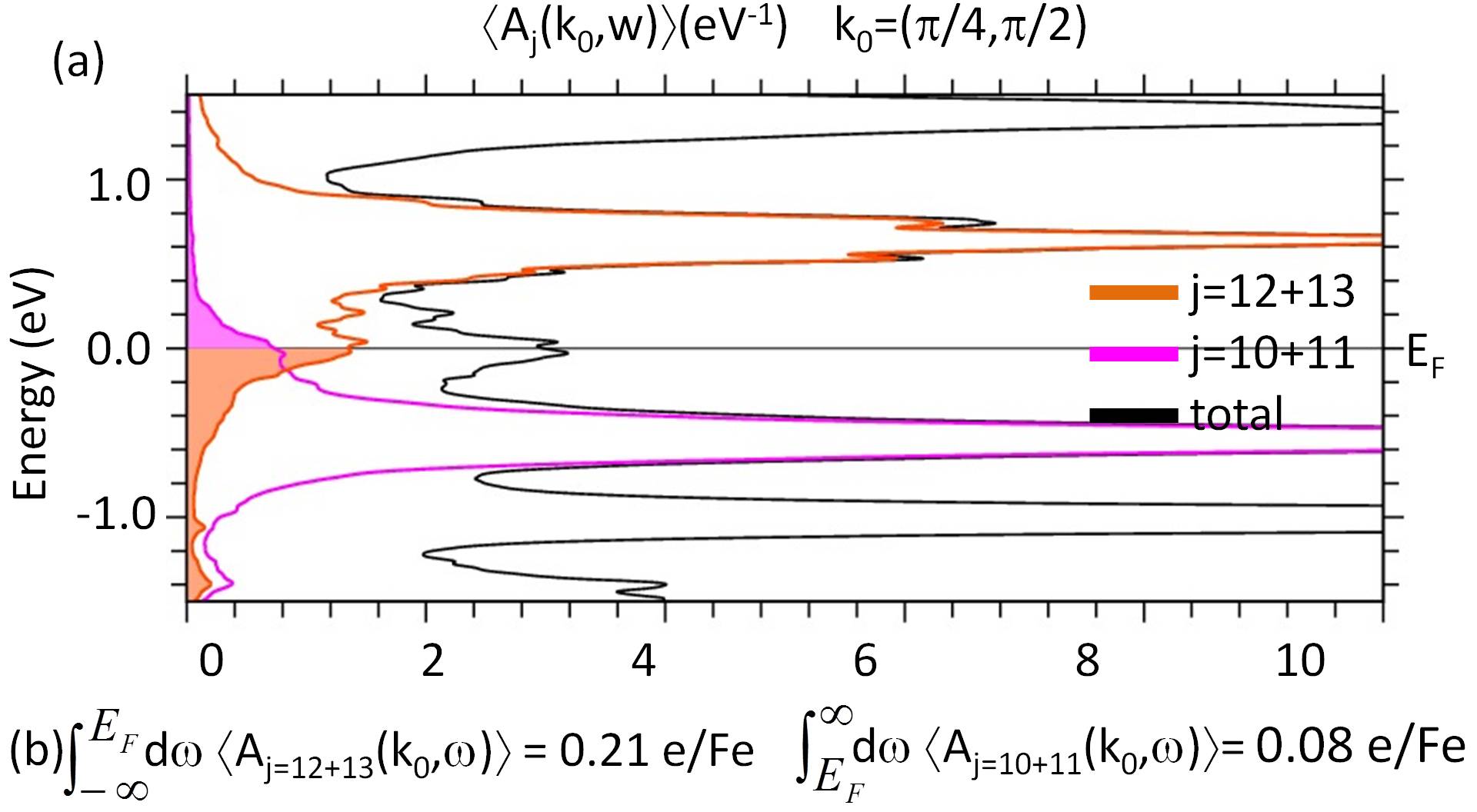}
\caption{\label{fig:fig3}
(color online) 
(a) Spectral function at fixed crystal momentum $k_0=(\pi/2,\pi/4)$ with resolved contributions from bands 10+11 and 12+13. (b) Integrated spectral weight below/above the Fermi energy of the bands 12+13/10+11. 
}\end{figure}

The disorder induced broadenings of the quasi-particle peaks in the Fe-$d$ bands are the reason why Se vacancies behave like hole dopants rather than weak electron dopants.  
The tails of the quasi-particle peaks pin the Fermi energy. 
Therefore if more quasi-particle weight leaks below the Fermi energy than above the Fermi energy will shift downward corresponding to an effective hole doping \cite{kxfeyse2}.
It will be instructive to illustrate this effect with the raw data of the current case of disordered FeSe$_{0.9}$. 
To this end we compute the band resolved spectral function $A_{j}(k,\omega)$ in the eigenbasis $|kj\rangle$ of the clean FeSe of crystal of momentum $k$ and band $j$.
As an example we show in Fig. \ref{fig:fig3}(a) the quasi-particle peaks of bands 10-13 in the energy range $[-1.5,1.5]$eV at crystal momentum $k_0=(\pi/2,\pi/4)$ corresponding to the bar indicated in the top panel of Fig. \ref{fig:fig2}(c).
The contributions from the degenerate bands 10 and 11 and the degenerate bands 12 and 13 have been added to better visualize the tails of the quasi-particle peaks.
From the numerical integration in \ref{fig:fig3}(b) we see that at this particular crystal momentum bands 12 and 13 leak more below the Fermi energy (0.21e/Fe) than bands 10 and 11 leak above (0.08e/Fe) resulting in a net contribution to the effective doping of 0.13 holes per Fe.
Of course the total effective doping is the sum of the contributions of all the bands at all the crystal momenta.  
This example illustrates how the disorder induced broadenings of the quasi-particle peaks control the Fermi energy and cause an effective doping of the bands.

Given that the Se vacancies behave like hole dopants rather than electron dopants, they cannot be the cause of the large electron doping in the FeSe monolayer as observed by ARPES ~\cite{dfliu, slhe, sytan}. 
This leaves open alternative scenarios such as the oxygen vacancy induced electron doping.
From ARPES measurements ~\cite{sytan} dispersionless spectral weight is found to be present in the SrTiO$_3$ substrate but absent after the FeSe monolayer is deposited.
This is interpreted as an electron transfer from oxygen vacancy states in the substrate to the Fe-$d$ bands in the FeSe monolayer.
However this does not explain what the crucial role of the post-annealing is.
Another interesting scenario ~\cite{mazin} is that during the Se etching oxygens are replaced with Se, which due to their large radius will induce Se vacancies in the FeSe monolayer. 
Building on this idea we can reverse the initial proposal. The post-annealing could move the Se from the substrate to the monolayer, thereby removing the Se vacancies rather then creating them. 
This would induce an electron doping as Se vacancies are removed and oxygen vacancies are created and would be consistent with the observation~\cite{sytan} that only the interfacial layer is doped but not the second monolayer.   
Furthermore this could explain why the Fermi surface becomes sharper during the annealing procedure since the oxygen vacancies will scatter the Fe-d bands less than the Se vacancies.
Obviously the study of such a complicated scenario would require the consideration of the substrate.

The unexpected doping effects of the anion vacancies reported in this study also have important implications for their controversial role in the iron based superconductors in general.
For example ARPES measurements ~\cite{zrye}  have shown an unexpected net hole doping in the isovalent substituted BaFe$_{2}$(As$_{1-x}$P$_x$)$_{2}$. 
But very puzzling, the same sample also displays a large amount ($\sim$10\%) of anion vacancies and/or excess Fe, which naively should have electron-doped the system strongly, instead.
A large VCFS occupancy in combination with an effective hole doping offers a most likely explanation of this anti-intuitive observation.

While some studies ~\cite{mcqueen} conclude that anion vacancies suppress the superconductivity, others~\cite{fchsu, nitsche,wli} conclude that they are an essential ingredient. 
To resolve this debate the microscopic effect of the anion vacancies on the Fe-$d$ carriers must be known. 
We have shown here that, counter-intuitively, the anion vacancies lower the Fermi energy which tunes the nesting conditions that govern the spin \cite{mazin2, kuroki, graser, chubukov} and orbital\cite{kontani} fluctuation theories of the superconductivity.   
Yet at the same time the anion vacancies only change the filling of the Fe-$d$ orbitals slightly since most of the doped electrons are absorbed by the VCFS orbital.
This means for example that anion vacancies will not significantly reduce the effect of the Hund's coupling which is known to be crucial for the correlations and the magnetism in these compounds~\cite{haule, zpyin}.
Furthermore, the very strong hybridization ($\sim$ eV) of the VCFS orbitals with their surrounding anion-$p$ orbitals will modify the strong coupling theories of the superconductivity\cite{kjseo, goswami}. 
A priory one would think that around the anion vacancy the magnetic exchange between the Fe-$d$ moments would be reduced since the hopping processes via the removed anions are deactivated. However, due to the surprising emergence of the VCFS state new types of hopping processes are allowed via the VCFS orbital. The derivation of these magnetic exchanges is an important task for future studies.
Finally the strong scattering by the anion vacancies as reflected by the large broadenings of the Fe-$d$ bands will allow the balance between the competing magnetism and superconductivity to be tipped~\cite{wadati,tm122, kikoin, hammerath, fernandes}.   
Together these important insights can shed new light on the controversial role of anion vacancies in the iron based superconductors. 

In conclusion, we investigated from first principles the doping effects of disordered Se vacancies on monolayer FeSe.
We find that Se vacancies pull a vacancy centered orbital below the Fermi energy that absorbs most of the doped electrons. 
Furthermore we find that the disorder induced broadening causes an effective hole doping. 
The surprising net result is that Se vacancies in terms of the Fe-$d$ bands behave as hole dopants rather than electron dopants. 
Our results exclude Se vacancies as the origin of the large electron pockets measured by ARPES.
The counter-intuitive doping effects found in this study also give new insight in the debated role of anion vacancies in the iron based superconductors, 
and exemplify the rich unexpected effects of vacancies in materials in general.

This work was supported by the U.S. Department 
of Energy, Basic Energy Sciences, Materials Sciences and
Engineering Division, and DOE-CMSN DE-AC02-98CH10886. 
PJH was supported by DOE DE-FG02-05ER46236.
HPC was supported by DOE DE-FG02-02ER45995.
TB was supported by DOE CMCSN and as a Wigner Fellow at the Oak Ridge National Laboratory.

\begin{widetext}
{\Large\bf Supplemental Material}

\section{Details of the density functional theory calculations}\label{sec:secdft}

We applied the WIEN2K\cite{blaha2} implementation of the full potential linearized augmented plane wave (LAPW) method in the local density approximation. 
The space group P4/nmm and the in-plane lattice constant $a=7.13$ Bohr of the monolayer Fe$_2$Se$_2$ normal cell were taken from Ref. ~\cite{margadonna2}. 
The out-of-plane lattice constant $c$ was chosen to be c=38.79 Bohr to include enough vacuum to treat the monolayer.  
The distance between the Fe and Se plane was found from a lattice relaxation to be z=0.062 (in units of the out-of-plane lattice constant $c$).
To capture the single Se vacancy influence, the $2\times 2$ Fe$_8$Se$_7$  supercell was used. 
The k-point mesh was taken to be 9$\times$9$\times$1 for the undoped normal cell and 4$\times$4$\times$1 for the supercell respectively.
The basis set sizes were determined by RKmax=7. 

\section{Details of the Wannier functions}\label{sec:secwan}

Given the Bloch states $|kj\rangle$ corresponding to a set of bands $\epsilon_{kj}$, where $k$ denotes the crystal momentum and $j$ denotes the band index, one can construct a set of Wannier states $|rn\rangle$ according to $|rn\rangle=\frac{1}{\sqrt{l}}\sum_{kj} e^{-ik\cdot r} |kj\rangle U_{jn}(k)$,
where $l$ denotes the number of unit cells in the system,  $r$ denotes the lattice vector and $n$ denotes the Wannier orbital index. 
The matrix $U_{jn}(k)$ fixes the so called gauge freedom of the Wannier functions.  
For this purpose we use the projected Wannier function method \cite{ku2, anisimov,thesis} in which this matrix is taken to be the projection of $n_{\rm{orb}}$ orbitals $|\varphi_{n} \rangle$ onto the Hilbert space of $n_{\rm{band}}(\geq n_{\rm{orb}})$ bands $|kj\rangle$. The Fe-$d$ and Se-$p$ Wannier functions are defined by projecting the corresponding atomic orbitals on the bands within the [-7,3]eV energy range. The VCFS Wannier function is defined by projecting the bonding molecular orbital of the four Fe-$s$ orbitals that surround the Se vacancy: $|\varphi_{\rm VCFS}\rangle=\sqrt{1/4}\left(|{\rm Fe1-}s\rangle+|{\rm Fe2-}s\rangle+|{\rm Fe3-}s\rangle+|{\rm Fe4-}s\rangle\right)$ 

\section{Vacancy centered Fe-s orbital for other supercells}\label{sec:secvcfs}

\begin{figure}[htp]
\includegraphics[width=1\columnwidth]{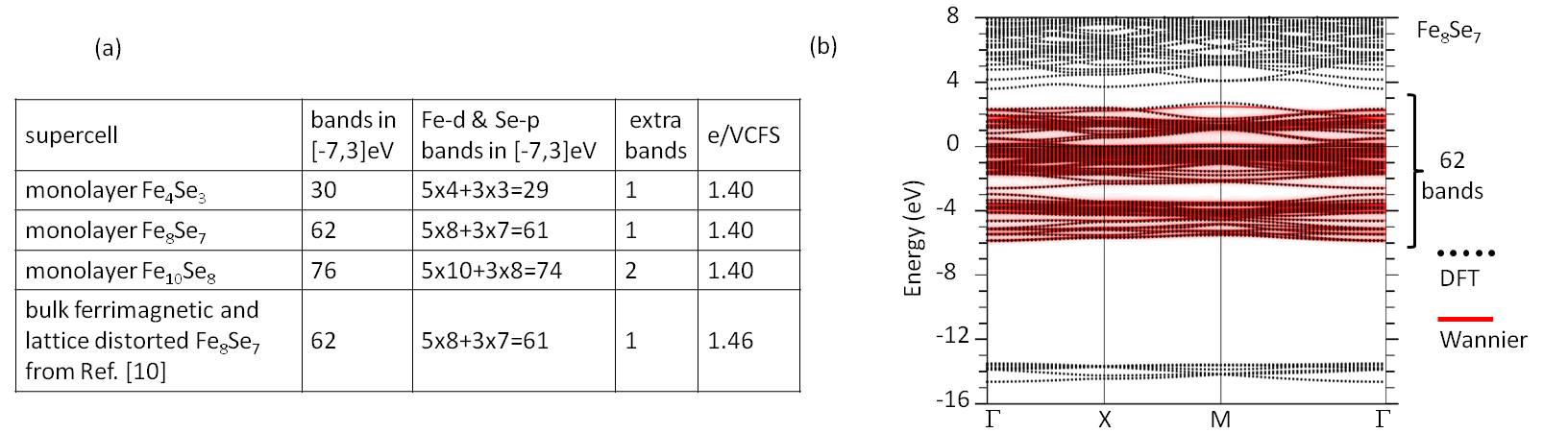}
\caption{\label{fig:figsup1}
(a) Number of vacancy centered Fe-$s$ (VCFS) bands and occupation for other supercells.
(b) Wannier and DFT bandstructure of the monolayer Fe$_8$Se$_7$ supercell.
}
\end{figure}

In Fig. \ref{fig:figsup3} the occupation analysis of the vacancy centered Fe-$s$ (VCFS) bands for other monolayer and bulk FeSe supercells is shown. The first three columns in Fig.  \ref{fig:figsup3}(a) show that per Se vacancy an additional band enters in the Fe-$d$/Se-$p$ band complex. The fourth column shows that its occupation is roughly $\sim$1.4 electrons per VCFS orbital. The last row shows that the conclusion holds when the Se vacancy induced ferrimagnetism and lattice relaxations are included that were reported earlier~\cite{kwlee2}. In Fig. \ref{fig:figsup3}(b) we show for the monolayer Fe$_8$Se$_7$ supercell the 62 bands within the [-7,3]eV energy range from which the $5\times8$ Fe-$d$,  the $3\times7$ Se-$p$ and the VCFS Wannier function are constructed.

\section{Benchmarks of the effective Hamiltonian against DFT}

To explore the accuracy and efficiency of the effective Hamiltonian method~\cite{naxco22} for the case of monolayer FeSe$_{1-x}$, we present comparisons of spectral functions $A_n(k,\omega)$  calculated from the full DFT and the effective Hamiltonian (Fig. \ref{fig:figsup2a} and \ref{fig:figsup2b}). The size of the deviations between the full DFT and the effective Hamiltonian should be compared with the size of the impurity induced changes. For this purpose the spectral function of the undoped Fe$_2$Se$_2$ is also plotted as a reference for each benchmark. The basis set of Linear Augmented Plane Waves (LAPW's) used in the full DFT is $\sim100$ times larger then the basis set of Wannier functions used in the effective Hamiltonian method. Since the number of floating point operations of diagonalization depends cubically on the size of the matrix this implies an efficiency increase by a factor of $100^{3}\approx10^{6}$. Furthermore the full DFT calculation involves multiple self consistent cycles ($\sim10$) whereas the effective Hamiltonian method requires only a single diagonalization, which increases the efficiency by another order of magnitude to $\sim 10^{7}$.
\begin{figure}[htp]
\includegraphics[width=0.8\columnwidth]{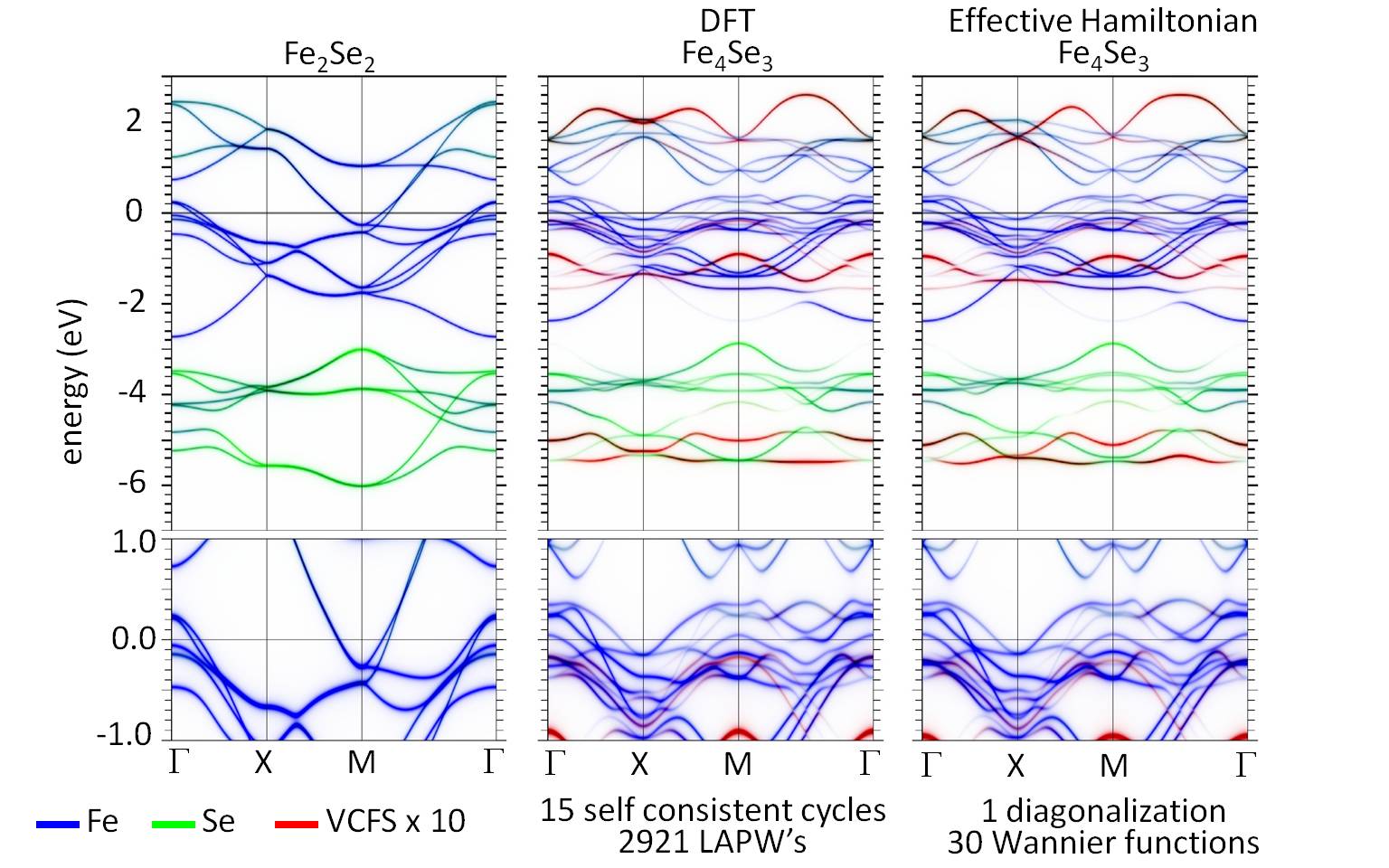}
\caption{\label{fig:figsup2a}
}
\end{figure}
\begin{figure}[htp]
\includegraphics[width=0.8\columnwidth]{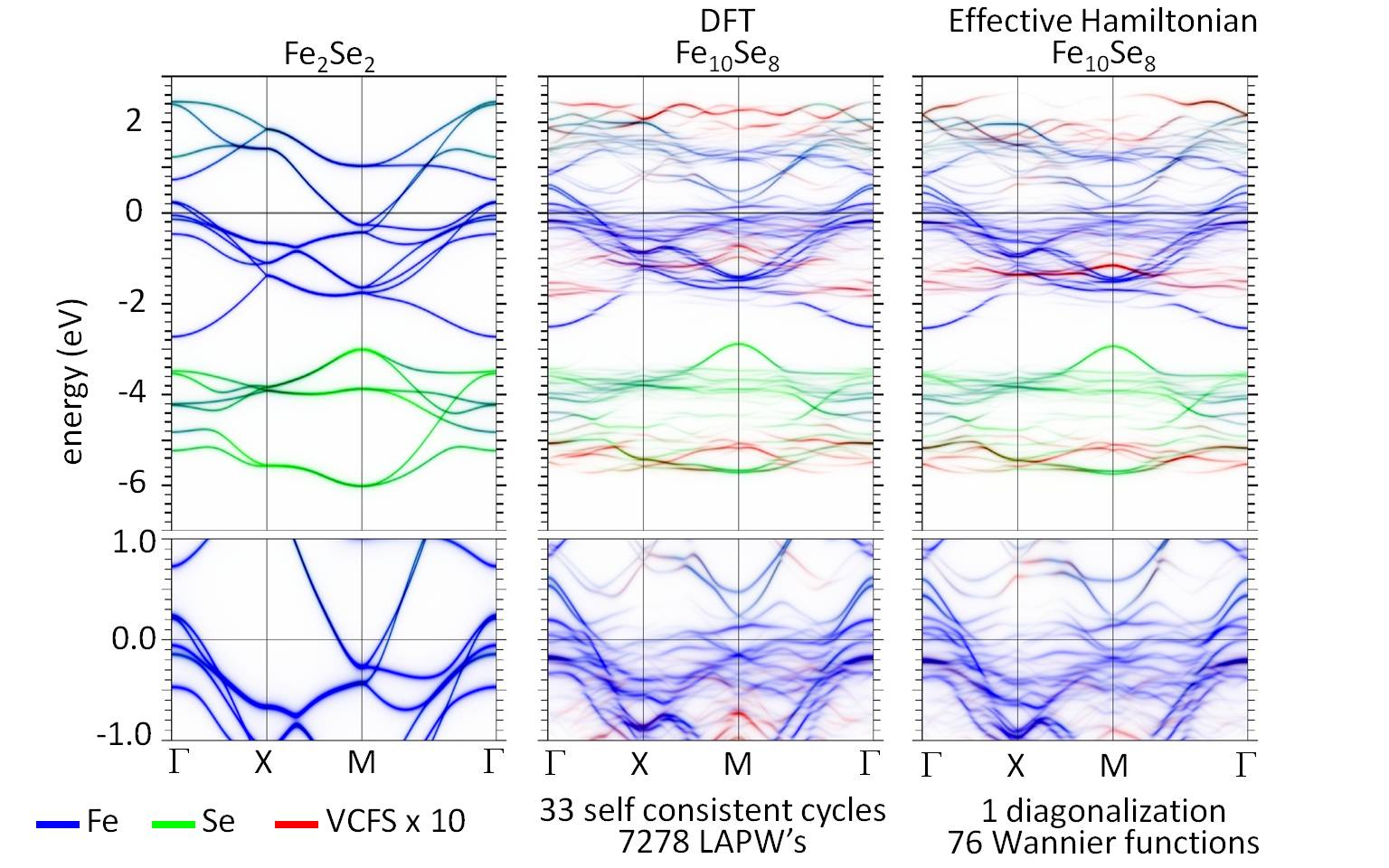}
\caption{\label{fig:figsup2b}
}
\end{figure}

\clearpage
\section{Convergence with respect to size and number of configurations}

\begin{figure}[htp]
\includegraphics[width=0.8\columnwidth]{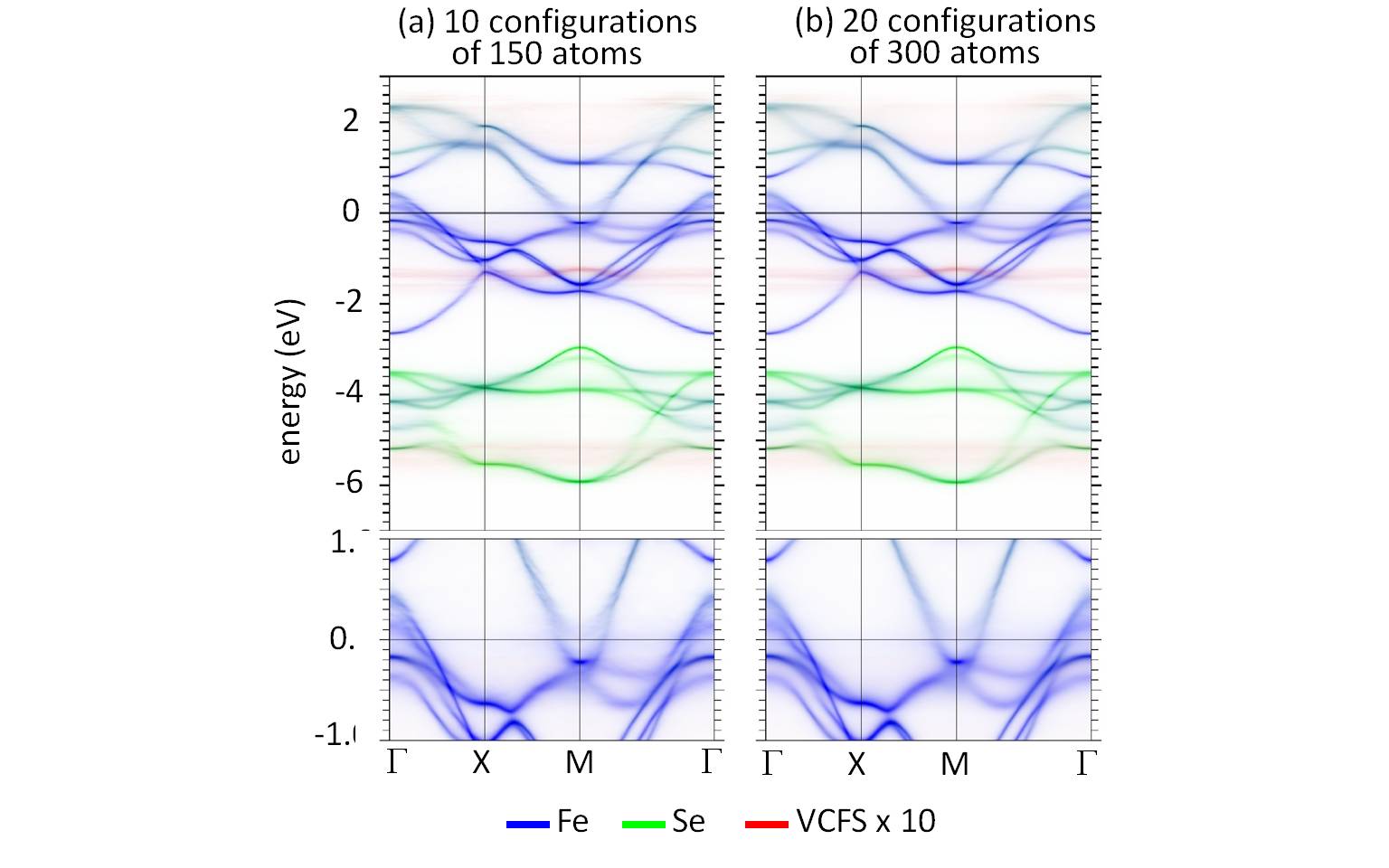}
\caption{\label{fig:figsup3} Bandstructures of disordered monolayer FeSe$_{0.9}$ from (a) 10 configurations 150 atoms on average and (b) from 20 configurations with 300 atoms on average.
}
\end{figure}

In Fig. \ref{fig:figsup3} we demonstrate the convergence of the spectral function of disordered FeSe$_{0.9}$ with respect to the number and the size of the configurations.

\end{widetext}

\end{document}